\documentclass[fleqn,10pt]{wlscirep}
\title{High power high repetition rate supercontinuum generation in graded-index multimode fibers}

\author[1,2]{U\u{g}ur Te\u{g}in}
\author[1,2,*]{B\"{u}lend Orta\c{c}}
\affil[1]{National Nanotechnology Research Center, Bilkent University, 06800 Bilkent, Ankara, Turkey}
\affil[2]{Institute of Materials Science and Nanotechnology, Bilkent University, 06800 Bilkent, Ankara, Turkey}

\affil[*]{ortac@unam.bilkent.edu.tr}

\begin{abstract}
Graded-index multimode fibers can handle high average powers and feature unique nonlinear dynamics due to spatio-temporal evolution of the light inside. We report on generation of octave-spanning spectrally flat supercontinua in graded-index multimode fibers using a compact, ytterbium-doped all-fiber laser system. Supercontinua with output powers as high as $\sim$4 W and 2 MHz repetition rate obtained with more than 1 $\mu$m spectral bandwidth. Evolution of supercontinuum is investigated by studying the effect of launch power, repetition rate, fiber length and core size. Numerical simulations are performed to investigate underlying nonlinear dynamics in details and calculated results are well-aligned with experimental spectra. To the best of our knowledge, this the highest average power and repetition supercontinuum source ever reported in a standard graded-index multimode silica fiber.\end{abstract}
\begin{document}

\flushbottom
\maketitle
%
%
\thispagestyle{empty}

\section*{Introduction}

Starting with standard silica fibers, supercontinuum generation in optical fibers is studied extensively in the past decades. Today fiber-based supercontinuum sources are largely used in a variety of applications including optical metrology, fiber communication systems and biomedical imaging \cite{alfano2016supercontinuum,dudley2010supercontinuum}. Evolution dynamics of supercontinuum generation inside of single-mode or few-mode fibers are well understood and photonic crystal fibers are allowed scientists to tailor key fiber properties to achieve octave spanning supercontinuum sources with a wide range of source parameters \cite{birks1997endlessly}. Recently graded-index multimode fibers (MMFs) drew attention by featuring unprecedented nonlinear phenomena. Intermodal interactions and spatio-temporal evolution with periodic imaging leads to discovery of unique phenomena such as Kerr self-cleaning \cite{krupa2017spatial,liu2016kerr}, spatio-temporal solitons \cite{renninger2013optical}, ultrabroadband dispersive waves \cite{wright2015ultrabroadband}, spatio-temporal instability \cite{longhi2003modulational,krupa2016observation,wright2016self,teugin2017observation}, quasi-phase matched four-wave mixing (FWM) \cite{nazemosadat2016phase} and cascaded Raman scattering \cite{pourbeyram2013stimulated}. 

With exploiting aforementioned nonlinear effects in graded-index MMFs a robust and relatively less sophisticated method to generate supercontinua is presented in the literature\cite{lopez2016visible,krupa2016spatiotemporal}. These efforts to generate supercontinuum in graded-index MMF by pumping in normal dispersion regime investigated the nonlinear dynamics by choosing low repetition rate (Hz-kHz) sources to achieve high pump peak powers. Hence maximum average powers of the reported supercontinua are in hundreds of mWs range so far even though graded-index MMFs are promoted with high power level handling potentials. Many applications of supercontinuum sources require achieving high average powers for detection purposes. All-fiber lasers are generally preferred to obtain high average and peak powers with high repetition rates. As a result of this capability, fiber based supercontinua are studied extensively in photonic crystal fibers with femtosecond to continuous-wave regimes \cite{dudley2006supercontinuum}. Hence fiber lasers are suitable supercontinuum source candidates to generate supercontinua in MMFs with high average powers. Here, we study the generation of octave-spanning high power and high repetition rate supercontinuum generation in graded-index MMFs with 62.5 $\mu$m core diameter (Thorlabs-GIF625). Pump pulses with MHz repetition rates, $\sim$30 kW peak power and 70 ps pulse duration are launched to graded-index MMF. Spectrally flat supercontinua with 1.88 W and 3.5 W output average powers are generated in 20 m fiber  with 1 MHz and 2 MHz repetition rates respectively. Although our system is pump power limited, our observations regarding power scaling with increasing repetition rate suggest >100 W powers can be achieved in graded-index MMFs with higher repetition rate pump laser systems.

\section*{Experimental Results}

\begin{figure}[ht!]
\centering
\includegraphics[width=1.0\textwidth]{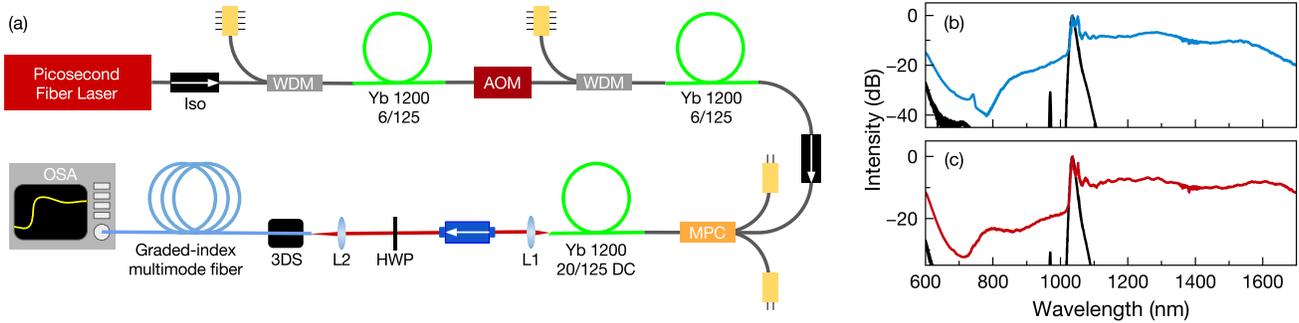}
\caption{(a) Schematic of the experimental setup comprising of Yb-doped fiber, wavelength division multiplier (WDM), isolator (Iso), acousto-optic modulator (AOM), multi-pump combiner (MPC), three-dimensional stage (3DS), half-wave plate (HWP). Spectra measured from (b) 10 m and (c) 20 m graded-index MMF for 1 MHz pump pulse repetition rate. Spectrum of pump pulse is presented as black.}
\label{fig:false-color}
\label{fig:1}
\end{figure}

Fig. \ref{fig:1} shows a schematic of the experimental setup. A home-built Yb-doped dispersion managed mode-locked fiber laser is employed as a pump pulse generator \cite{ortacc200390}. Chirped pulses are first amplified by a preamplifier to $\sim$70 mW average power. AOM is employed to change fundamental repetition rate of the pulse train before the main amplifier. Due to the loss of AOM and the change of repetition rate from $\sim$40 MHz to 1 MHz average power drops to $\sim$ 2  mW after the AOM. Thus another preamplifier is placed before the double clad main amplifier. At the end of the main amplifier, we obtain 70 ps pulses centered around 1045 nm with $\sim$20 nm bandwidth and adjustable repetition rates (kHz-MHz) which corresponds to $\sim$30 kW peak power. We collimate the output of the system with a biconvex lens and high power free-space isolator is used to prevent back reflections. We use a half-wave plate to change the polarization of the pump pulse to achieve best conversion condition in graded-index MMF. To excite the fiber we use a biconvex lens with 2 cm focal length which creates $\sim$ 20 $\mu$m beam waist size at the facet of the MMF. Optimum coupling condition is achieved with a three-axis translation stage which enables free space coupling efficiency greater than 80\%. 

By using pump pulses with 1 MHz repetition rate generation of supercontinuum is demonstrated in 20 m and 10 m graded-index MMFs with 2.19 W and 1.88 W output average power as presented in Fig. \ref{fig:1}(b-c). Pump pulse spectrum contains low unabsorbed pump around $\sim$ 980 nm but at the end of graded-index MMF, supercontinuum become dominated in that region as well. Spectral measurements are performed with an optical spectrum analyzer covering a range from 600 nm to 1700 nm. Due to the lack of a suitable measurement tool, no information above 1700 nm obtained thus end of generated supercontinua on the longer wavelengths is unknown. By comparing results of 10 m and 20 m fiber we notice increasing the propagation length helps to generate conversions for shorter wavelengths and broader spectrum can be achieved but due to the losses of fiber average output power starts to drop. 

\begin{figure}[ht!]
\centering
\includegraphics[width=1.0\textwidth]{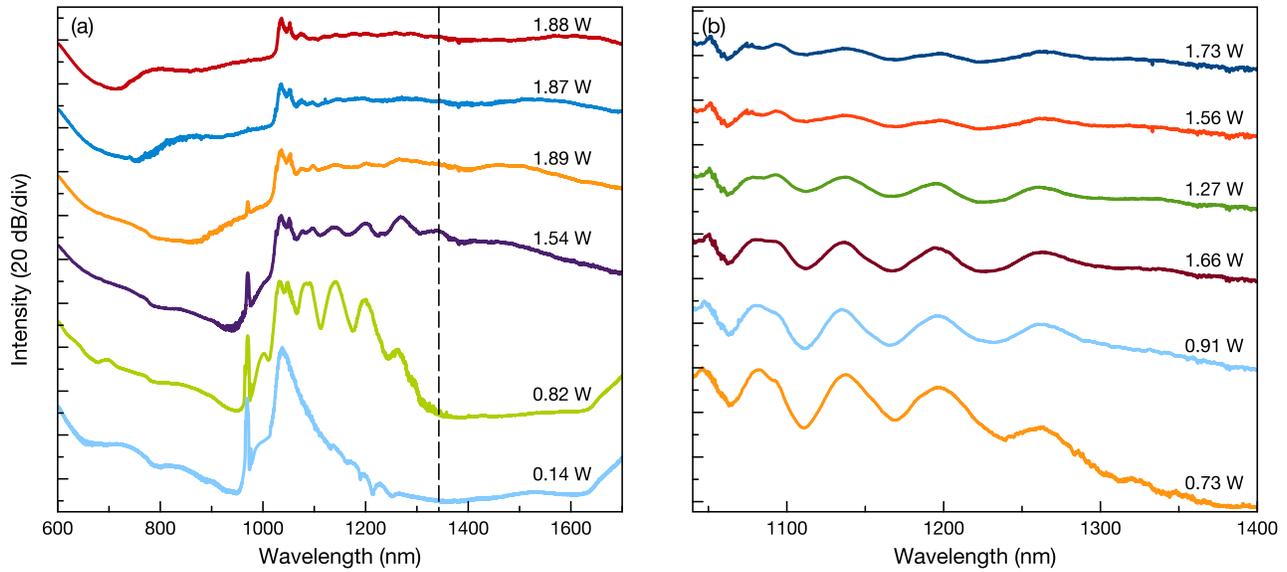}
\caption{Evolution of (a) supercontinuum spectrum and (b) SRS peaks inside 20 m graded-index MMF as a function of launched pulse average power recorded for 1 MHz pump pulse repetition rate. Output average powers indicated for each spectrum.}
\label{fig:Fig2}
\end{figure}

Evolution of supercontinuum for 20 m fiber is investigated in detail by studying the effect of pump launch power [Fig. \ref{fig:Fig2}]. For low powers, we observe small wavelength conversion from pump to 780 nm and 1550 nm. These frequency shifts are matching with intermodal FWM and spatiotemporal instability but with increasing launch power SRS becomes dominant nonlinear mechanism. When 0.82 W output power is reached, generation of cascaded SRS with $\sim$ 13 THz frequency shifts is observed as reported by Pourbeyram et al. \cite{pourbeyram2013stimulated} [Fig. \ref{fig:Fig2}(b)]. When SRS peaks reach zero dispersion wavelength (ZDW) of the fiber ($\sim$ 1330 nm), broad spectral formation is emerges from the noise. This behavior can be explained by complex parametric phenomena including collision-based spectral broadening and the reduction in SRS gain near ZDW \cite{dudley2010supercontinuum,golovchenko1990mutual,vanholsbeeck2003complete}. After we reached 1.89 W output power we observed a sudden drop at average output power and wavelength generation at shorter wavelengths starts to emerge. Evolution of supercontinuum at visible wavelengths can be explained with the coupling of Raman and parametric gain which also takes place in supercontinuum generation for picosecond pulses in photonic crystal fibers \cite{dudley2006supercontinuum}. This results in the generation of anti-Stokes wavelengths even without proper phase-matching \cite{coen2002observation}. In the end, more than octave spanning spectrally flat supercontinuum achieved with 1.88 W average output power for 1 MHz pump repetition rate. Overall spectral intensity deviation of the continuum is calculated as 52\%. On the other hand, above the pump wavelength (between 1060-1700 nm), spectral intensity deviation is incredibly small as 24\%.

\begin{figure}[b!]
\centering
\includegraphics[width=0.95\textwidth]{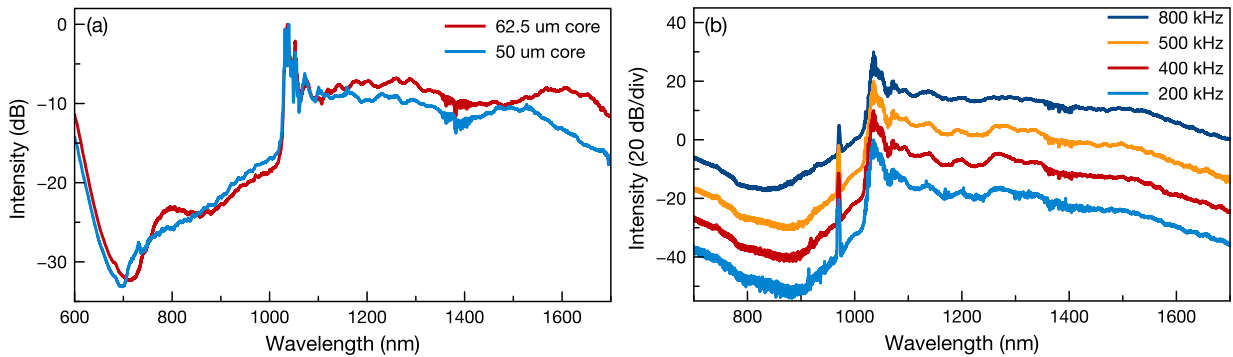}
\caption{(a) Supercontinuum spectra measured from 20 m graded-index MMF (62.5 $\mu$m core diameter) 200 kHz to 800 kHz repetition rates for constant peak power. (b) Supercontinuum spectra for graded-index MMFs with different core diameters.}
\label{fig:Fig5}
\end{figure}

We study supercontinuum generation in 20 m graded-index multimode fiber with 50 $\mu$m core diameter to demonstrate versatility of this low-cost supercontinuum generation method. For 1 MHz repetition rate and same output average power, spectral difference is presented in Fig. \ref{fig:Fig5}(a). Spectral features resembles the supercontinuum generated in graded-index multimode fiber with 62.5 $\mu$m core diameter. To demonstrate power scalability of the supercontinuum generation method in graded-index MMF, while peak power of the pump pulse remains same as we sweep repetition rate of pump pulses in kHz region [Fig. \ref{fig:Fig5}(b)]. In the experiments we set the pump peak power as 25 kW and increase average output power from 350 mW to 1.4 W. 

\begin{figure}[ht!]
\centering
\includegraphics[width=0.97\textwidth]{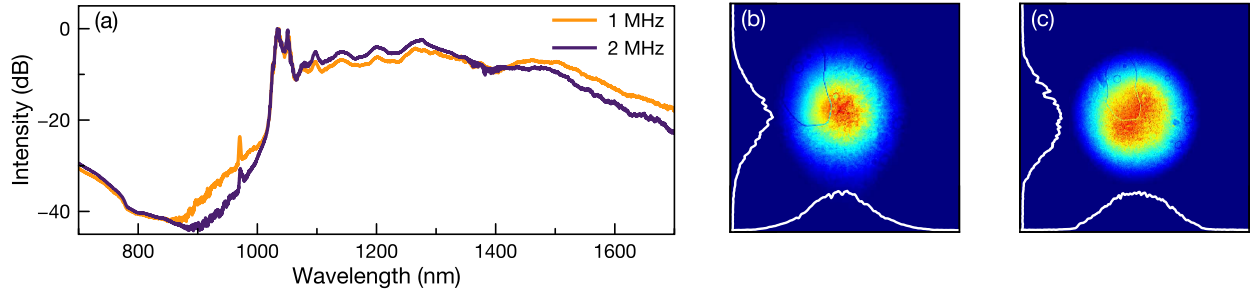}
\caption{(a) Supercontinuum spectra for MHz repetition rates with same peak power measured from  20 m graded-index MMF. Near-field beam profile for (b) 730 nm to 1200 nm range and (c) 1100 nm to 1200 nm range of the supercontinuum.}
\label{fig:Fig4}
\end{figure}

Here we focus on MHz repetition rates and as shown in Fig. \ref{fig:Fig4}(a), with increasing pump pulse repetition rate from 1 MHz to 2 MHz by preserving pump peak power ultra-broad supercontinua could be reproduced. By doubling pump pulse repetition rate from 1 MHz to 2 MHz, while keeping peak powers constant, we achieve 3.96 W and 3.50 W average powers for supercontinua generated in 10 m and 20 m graded-index MMF, respectively. Our scaling experiments are pump power limited but these results indicate that by increasing average power and repetition rate, a hundred watts could be obtained with standard graded-index fiber while octave-spanning supercontinuum features remain.

Next, we measure near field spatial distribution of supercontinuum using a beam profiler operating up to 1200 nm. Due to the device limitation, we could measure the beam profile from 730 nm to 1200 nm Fig. \ref{fig:Fig4}(b). Additionally with a long-pass filter, we measure beam profile from 1100 nm to 1200 nm Fig. \ref{fig:Fig4}(c), as well. Even though fiber preferred in the experiments supports hundreds of modes, Gaussian-like spatial profile with high-order modes in the background is observed. Raman or Kerr beam cleaning could be the reason of observed spatial distributions \cite{krupa2017spatial,terry2007explanation}. 

\subsection*{Numerical Results}

In order to develop better understanding of supercontinuum evolution inside graded-index MMF we perform numerical simulations with 1+1D generalized nonlinear Schr\"{o}dinger equation: 
\begin{equation}
\frac{\partial A}{\partial z} +\left ( \sum_{n\geq 2}\beta _{n}\frac{i^{n-1}}{n!}\frac{\partial^n }{\partial t^n} \right )A=i\gamma (z)\left (  1+\frac{\partial }{\partial t}\right ) \left ( (1-f_{R})A\left | A \right |^{2}+f_{R}A\int_{0}^{\infty}h_{R}(t')\left | A(z,t-t') \right |^{2}\right )
\label{eq:refname2}
\end{equation}
with a periodic nonlinear coefficient \cite{wright2015ultrabroadband,conforti2017fast}. For numerical integration with high accuracy in simulations, we prefer the fourth-order Runge-Kutta in the Interaction Picture (RK4IP) method \cite{hult2007fourth,csenel201333}. In simulations, we include Raman process ($f_{R}$), shock terms and high-order dispersion coefficients up to $\beta _{7}$. SRS is included in the equation via use of a response function \cite{stolen1989raman}. Even though our simulations start from quantum noise, we average the simulations over 20 sets of initial conditions to simulate experimental observations more accurately. To decrease computation time, we consider pump pulse duration 10 times smaller than the experiments while pulse energy remains the same. It allows us to decrease propagation length 10 times. Therefore simulations are performed with 7 ps pulse duration, $\sim$300 kW peak power at 1040 nm central wavelength and we set $n_{0}$ as 1.470, $n_{2}$ as $2.7x10^{-20} m^{2}/W$, relative index difference as $0.01$, time window width as 70 ps with 2 fs resolution.

\begin{figure}[ht!]
\centering
\includegraphics[width=0.97\textwidth]{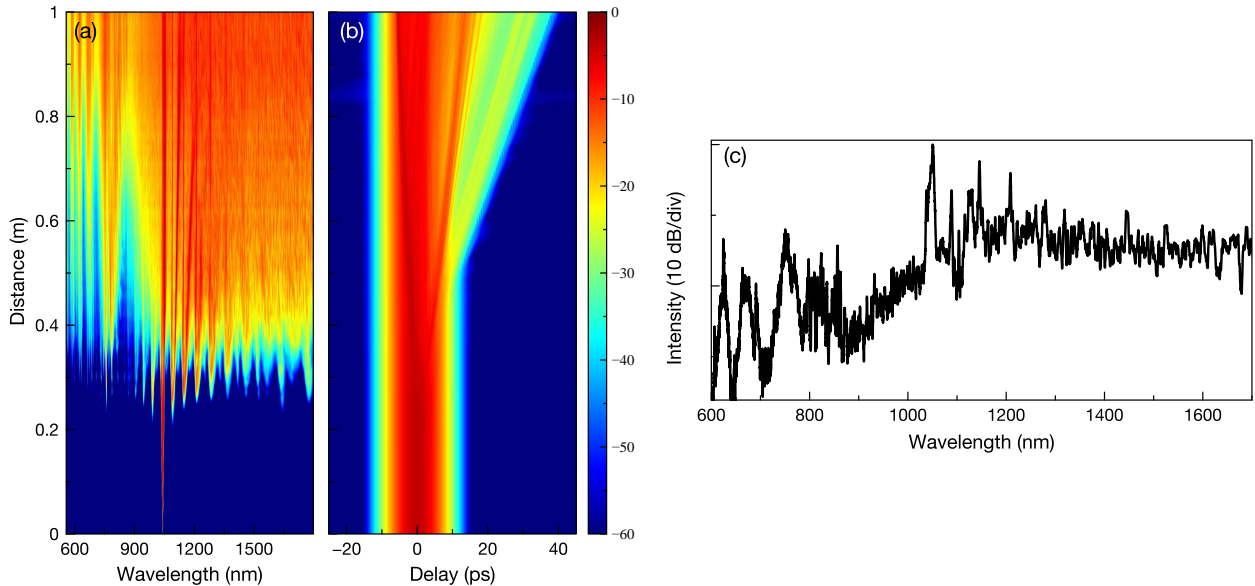}
\caption{Results obtained by averaging of numerical simulations showing (a) spectral and (b) temporal evolution through 1 m graded-index MMF with 62.5 $\mu$m core diameter. The intensities are in dB scale. Numerically obtained (c) final spectrum at the end of the fiber.}
\label{fig:Fig3}
\end{figure}

Spectral and temporal evolution obtained from numerical simulations for 1 m graded-index MMF with 62.5 $\mu$m core diameter are presented in Fig. \ref{fig:Fig3}. Obtained results are equivalent to experimental supercontinuum generation after propagation of 10 m fiber. Detailed analysis on the evolution of the pump pulse inside the fiber reveals that simulations also present strong cascaded SRS generation with relatively low parametric wavelength conversions in the spectral domain. After SRS peaks reach to ZDW, generation of new wavelengths at anomalous dispersion resembles development of Raman soliton components. In the literature, this phenomenon is explained as the transformation of SRS peaks above ZDW to ultrashort solitons \cite{golovchenko1991numerical}. These solitons experience self-frequency shift and more uniform spectra can be formed. Preceding soliton dynamics lead to temporal breakup seen in the simulations after 0.5 m propagation and with the help of spectrally broadened SRS peaks, a supercontinuum starts to appear. Spectral flatness and intensity distribution behavior matches well with experiments even though loss terms are not included in the simulations. 

\section*{Conclusion}

In summary, we demonstrated the generation of high average power spectrally flat octave-spanning supercontinua using a graded-index MMF pumped with an all-fiber laser system. The highest supercontinuum output power of 3.96 W is achieved in graded-index MMF with 62.5 $\mu$m core diameter using picosecond pulses at MHz repetition rate. Numerical simulations reveal that unique cascaded SRS observed in graded-index MMF plays a significant role in the spectral evolution. We have shown that this recently discovered, low-cost graded-index multimode fiber based supercontinuum source could benefit from multimode features of the fiber and high power level average powers are feasible to achieve. Adaptability of the high power and high repetition rate supercontinuum generation method is studied as well. Further power scaling based on fiber pump lasers enable high repetition rate all-fiber supercontinuum systems with >100W average powers. Therefore noise and coherency measurements of the supercontinua generated with graded-index MMF studies become feasible.

\section*{Acknowledgments}
The authors thank TUBITAK, TUBA-GEBIP, BAGEP, METU Prof. Dr. Mustafa Parlar Foundation and FABED for support and {\c{C}}. {\c{S}}enel for insightful helpful suggestions.

\bibliography{sample}

\end{document}